\documentclass[aps,pre,superscriptaddress,twocolumn,showpacs,floatfix]{revtex4}

\usepackage{amsfonts}
\usepackage{amssymb}
\usepackage{epsfig}
\usepackage{graphicx}
\usepackage{subfigure}

\newcommand{\uo}[1]{\underline{\omega}_{#1}}

\begin{document}
\title{Fractality of the non-equilibrium stationary states of open\\
  volume-preserving systems: I. Tagged particle diffusion} 
\author{Felipe Barra}
\affiliation{Departamento de F\'{\i}sica, Facultad de Ciencias
F\'{\i}sicas y Matem\'aticas, Universidad de Chile, Casilla 487-3,
Santiago Chile}
\author{Pierre Gaspard}
\affiliation{Center for Nonlinear Phenomena and Complex Systems,
  Universit\'e Libre  de Bruxelles, C.~P.~231, Campus Plaine, B-1050
  Brussels, Belgium}
\author{Thomas Gilbert}
\affiliation{Center for Nonlinear Phenomena and Complex Systems,
  Universit\'e Libre  de Bruxelles, C.~P.~231, Campus Plaine, B-1050
  Brussels, Belgium}
\date{\today}
\begin{abstract}
Deterministic diffusive systems such as the periodic Lorentz gas,
multi-baker map, as well as spatially periodic systems of interacting
particles, have non-equilibrium stationary states with fractal 
properties when put in contact with particle reservoirs at their
boundaries. We study the macroscopic limits of these systems and
establish a correspondence between the thermodynamics of the
macroscopic diffusion process and the fractality of the stationary
states that characterize the phase-space statistics. In particular the
entropy production rate is recovered from first principles using a
formalism due to Gaspard  [J. Stat. Phys. \textbf{88}, 1215 (1997)]. This
article is the first of two; the second article considers the influence of
a uniform external field on such systems. 
\end{abstract}
\pacs{05.45.-a,05.70.Ln,05.60.-k}
\maketitle
\section{\label{sec.int}Introduction}

The assumption that the microscopic dynamics of mechanical systems obeying
Newton's equations are mixing offers a mechanism by which the entropy can
increase toward its equilibrium value, as described by Gibbs in 1902
\cite{Gibbs}. The mixing would indeed allow coarse-grained probabilities to
reach their equilibrium values after a long time, a result that has
received a rigorous meaning in the context of modern ergodic theory
\cite{CFS82}. The use of the coarse-grained entropy of a physical
system, which is the second ingredient of Gibbs' mechanism, is
justified by the fact that, if the entropy should be given according to
Boltzmann by the logarithm of the number of complexions of a system,
then the introduction of cells of non-vanishing sizes is required to
perform the counting of complexions in systems described by continuous
coordinates. 

The program set up by Gibbs has taken on a new perspective in recent years
with its systematic application to chaotic, deterministic volume-preserving
dynamical systems sustaining a transport process of 
diffusion, which therefore fall under the scope of Liouville's
theorem \cite{TG95, Gas97a, TG99, TG00, GD00, GDG00, DGG02}. In previous 
works, both non-equilibrium stationary states and relaxation to equilibrium
were considered, with the common thread that transport processes are
intimately connected to singularities of the non-equilibrium measures,
characterized by phase-space distributions with fractal
properties.

A specific example of the systems of interest is the periodic Lorentz gas,
in which moving particles diffuse through a lattice interacting only with
fixed scatterers. A useful simplification of this process is the
multi-baker map, 
which is a simple model of deterministic diffusion for a tracer particle
with chaotic dynamics. Other higher dimensional examples are spatially
periodic extensions of systems with many interacting particles, where the
motion of 
a tagged particle is followed as it undergoes diffusion among the cells. 

In refs. \cite{GDG00, DGG02}, these models were studied in the
context of relaxation to equilibrium. It was shown in these papers that the
initial non-equilibrium distribution function rapidly develops a fractal
structure in phase space due to the chaotic nature of the dynamics. This
stucture is such that variations of the distribution function on
arbitrarily fine scales grow as the system evolves in time. The final
stages of the approach to equilibrium are then controlled by the decay of
fractal, microscopic hydrodynamic modes of the system --in this case
diffusive modes-- which 
decay with time as $\exp(-\mathcal{D} k^2 t)$, where $k$ is a wave number
associated to the macroscopic hydrodynamic decay, $\mathcal{D}$ is the
diffusion coefficient, and $t$ is the time. It is possible to express the
rate of entropy production in this final stage in terms of measures
which are determined by the non-equilibrium phase-space distribution,
specified by the fractal hydrodynamic modes. The main result is that one
obtains by this method exactly the 
expression for the rate of entropy production as given by irreversible
thermodynamics for these systems \cite{deGMaz}. The source of this agreement
can be traced  to the role played by the fractal hydrodynamic modes, both
for requiring a coarse graining of the phase space to properly incorporate
the effects of their fractal properties on entropy production in the
system, as well as for describing the slowest decay of the system as it
relaxes to equilibrium.

The purpose of this paper is to consider in further details the
non-equilibrium stationary states of multi-baker maps and Lorentz gases
which occur when their boundaries are put in contact with
particle reservoirs, as well as to extend these considerations to tagged
particle diffusion in spatially periodic systems of interacting particles.
In the macroscopic limit, we establish the connections between the
statistics of these deterministic systems and the phenomenological
prescriptions of thermodynamic. Though the stationary states of multi-baker
maps and Lorentz gases have been considered in some details, see especially
\cite{Gas98}, the problem of computing the entropy production associated to
the non-equilibrium stationary state has thus far been limited to the
example of the multi-baker map \cite{Gas97a}. Here we emphasize the
similarities between the stationary states of multi-baker maps, Lorentz
gases, and spatially periodic many particle systems, undergoing steady mass
flows and show how the formalism described in \cite{DGG02} yields an
\emph{ab initio} derivation of the entropy production rate in all these
systems.

This paper is the first of two. In the second paper \cite{BGGii}, we will
consider the Galton board \cite{Galton1889, mathworld}, which is to be
understood in this context as a periodic finite horizon Lorentz gas with a
uniform external field and no dissipation mechanism. As was recently proved
by Chernov and Dolgopyat \cite{CD07a, CD07b}, this system is recurrent and
therefore has no drift. We will show how one can characterize equilibrium
and non-equilibrium stationary states of such a system, much in the same
way as with the periodic Lorentz gas we consider in this paper.

The plan of the paper is as follows. In Sec. \ref{sec.phe}, we review the
phenomenology of non-equilibrium diffusive systems and their entropy
production. Then, starting from the Liouvillian evolution of phase-space
distributions, we construct the non-equilibrium stationary states of
open periodic Lorentz gases in Sec.~\ref{sec.olg}, as well as of
multi-baker maps in Sec. \ref{sec.mbm}, and show how such stationary states
naturally separate between regular and singular parts. While the regular
parts have the form of a local equilibrium and
allow, in the continuum limit, to retrieve the phenomenological solution
of a diffusive system undergoing a mass flow, the singular parts have no
macroscopic counterpart and encode the dynamical details of the systems'
diffusive properties. In Sec. \ref{sec.sps}, these results are extended to
systems of many interacting particles with non-equilibrium flux boundary
conditions. Under the assumption that the dynamics of these systems 
is chaotic, their stationary states are shown to have properties similar to
those of multi-baker maps and Lorentz gases. The singular part of the
stationary states is at the origin of the entropy production rate, as is
explained in Sec. \ref{sec.ep}, where the phenomenological entropy
production rate is retrieved from \emph{ab initio}
considerations. Conclusions are drawn in Sec. \ref{sec.con}.

\section{\label{sec.phe}Phenomenology}

The thermodynamics of deterministic models of diffusion such as the Lorentz
gas and multi-baker map to be discussed below was extensively reviewed by
Gaspard \emph{et.~al} in \cite{GND03}. In these models, independent tracer
particles mimic matter exchange processes of binary mixtures, \emph{i.e.}
the process of mutual diffusion between the light tracer particles and the
infinitely heavy particles that constitute the background. 

In such systems, the irreversible production of entropy arises from
gradients in the density of tracer particles. For a dilute system, the
resulting thermodynamic force is 
\begin{equation}
  \vec{\mathcal{X}}(\vec{X},t) = - \frac{\vec{\nabla}
    \mathcal{P}(\vec{X},t)} {\mathcal{P}(\vec{X},t)}\,,
  \label{thermoforce}
\end{equation}
where $\mathcal{P}(\vec{X},t)$ denotes the density of
tracer particles at position $\vec{X}$ and time $t$. Here the 
Boltzmann constant is set to unity. Close to equilibrium, in the linear
range of irreversible processes, the corresponding current is proportional
to the thermodynamic force and given by Fick's law of diffusion according
to 
\begin{equation}
  \vec{\mathcal{J}}(\vec{X},t) = -\mathcal{D} \vec{\nabla}
  \mathcal{P}(\vec{X},t)\,, 
  \label{thermocurrent}
\end{equation}
where $\mathcal{D}$ is the coefficient of diffusion of tracer
particles (assumed to be uniform). In this limit, the condition of mass
conservation, $\partial_t \mathcal{P}(\vec{X},t) +
\vec{\nabla}\cdot\vec{\mathcal{J}}(\vec{X},t) 
= 0$, transposes into the the Fokker-Planck equation,
\begin{equation}
  \partial_t \mathcal{P}(\vec{X},t) = \mathcal{D} \nabla^2 \mathcal{P}
  (\vec{X},t)\,.
  \label{FPeq}
\end{equation}

The product of the thermodynamic  force and current yields the
rate of local irreversible production of entropy,
\begin{eqnarray}
  \frac{d_\mathrm{i} \mathcal{S}(\vec{X},t) }{dt}
  &=& \vec{\mathcal{J}}(\vec{X},t)\cdot
  \vec{\mathcal{X}}(\vec{X},t)\,,\nonumber\\
  &=& \mathcal{D} \frac{[\vec{\nabla} \mathcal{P}(\vec{X},t) ]^2}
  {\mathcal{P}(\vec{X},t) } > 0\,.
  \label{thermoep}
\end{eqnarray}

Our goal in the next sections will be to analyze the statistics of
deterministic models of diffusion, identify the conditions under which
their statistical evolution reduces to the evolution of macroscopic
densities as described by Eq.~(\ref{FPeq}) and analyze the microscopic
origins of the production of entropy prescribed according to
Eq.~(\ref{thermoep}), which manifest themselves in the fractality of the
stationary states of these models. 

\section{\label{sec.olg}Open periodic Lorentz gas}

\begin{figure*}[thb]
  \centering
  \includegraphics[width=\textwidth]{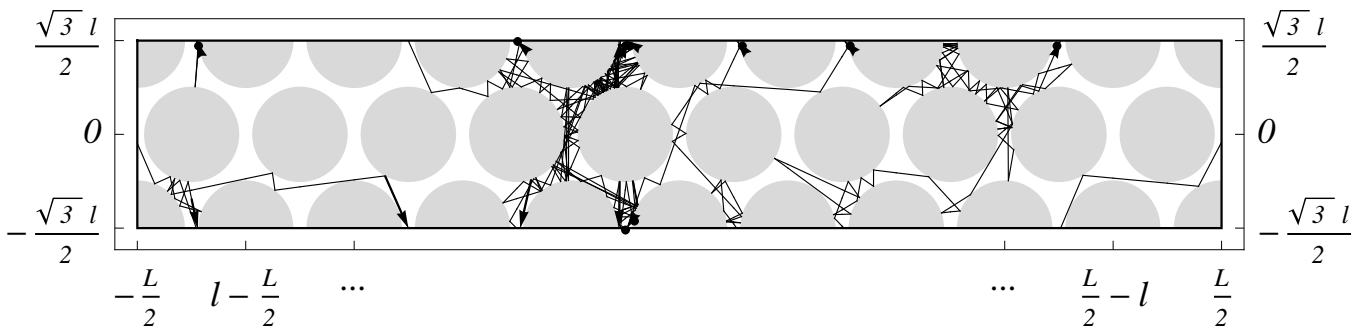}
  \caption{Cylindrical Lorentz channel with periodic boundary conditions at
    the upper and lower borders. The length of the channel is $L$, with $N$
    identical cells of widths $l$ and heights $\sqrt{3}l/2$, $N l \equiv 
    L$, each containing two disks. The disks have radii $\sigma = 0.44l$,
    and are positioned at points $(x,y) = ( nl/2, 0)$, $n =
    -N+1,-N+3,\dots,N-1$,  and $(n l/2, \pm \sqrt{3}l/2)$, $n =
    -N,-N+2,\dots,N-2,N$. The trajectory displayed is that of a particle
    which initially starts at unit velocity from the left-hand boundary,
    and is eventually absorbed as it reaches the right-hand
    boundary. Arrows on the upper and lower borders indicate points where
    the trajectory winds around the cylinder.
  }
  \label{fig.lg}
\end{figure*}

In this section, we consider a two-dimensional finite horizon periodic
Lorentz gas with hexagonal symmetry, in the shape of a cylindrical Lorentz
channel of length $L$, with periodic boundary conditions in the
$y$-direction and absorbing boundaries at $x = \pm L/2$. Though 
some of the results presented in this section are original, for the most
part this material is a review of existing results that can be found in
Ref. \cite{Gas98} and references therein. 
  
The geometry of the channel is displayed in Fig.~\ref{fig.lg} with a
typical trajectory. The channel consists
of a cylinder of length $L = N l$ and height $\sqrt{3}l$ with disks
$\mathbb{D}_n$, $-N\leq n\leq N$, of radii $\sigma$,
$\sqrt{3}/4<\sigma/l<1/2$. Assuming $N$ even, the disks' centers take
positions 
\begin{equation}
  (x_n,y_n) = 
  \left\{
    \begin{array}{l@{\quad}l}
      (nl/2, 0)\,,&n\,\mathrm{odd},\\
      (n l/2, \pm \sqrt{3}l/2)\,,& n\,\mathrm{even},
    \end{array}
  \right.
\end{equation}
where the disks at $y = \pm \sqrt{3}l/2$ are identified. We will also
denote by $\mathbb{I}_n$ the cylinder region around each disk $\mathbb{D}_n$,
\begin{equation}
  \mathbb{I}_n = \big\{(x,y)\ |\ (n - 1/2)l/2 \leq x \leq (n + 1/2)l/2
  \big\}.
\end{equation}
Thus the interior of the cylinder, where particles propagate freely is made
up of the union $\cup_{n = -N}^N \mathbb{I}_n \setminus \mathbb{D}_n$.

The associated phase space, defined on a constant energy shell, is
$\mathbb{C} = \cup_{n=-N}^N \mathbb{C}_n$, where $\mathbb{C}_n =
\mathbb{S}^1 \otimes [\mathbb{I}_n \setminus \mathbb{D}_n]$ and the unit
circle $\mathbb{S}^1$ represents all the possible velocity
directions. Particles are reflected with  elastic collision rules on the
border $\partial \mathbb{C}$, except at the external borders, corresponding
to $x = \pm L/2$, where they get absorbed. 
Points in phase space are denoted by $\Gamma$, $\Gamma = (x,y,v_x,v_y)$,
with fixed energy $E \equiv (v_x^2 + v_y^2)/2$, and trajectories by 
$\Phi^t\Gamma$, with $\Phi^t$ the flow associated to the dynamics
of the Lorentz channel. 

The collision map takes the point $\Gamma = (x, y,
v_x, v_y) \in \partial \mathbb{C}$ to $\Phi^\tau \Gamma = (x',y',v_x',v_y')
\in \partial \mathbb{C}$, where $\tau$ 
is the time that separates the two successive collisions with the border
of the Lorentz channel $\partial \mathbb{C}$, and $(v_x',v_y')$ is obtained
from $(v_x, v_y)$ by the usual rules of specular collisions. Given
that the energy $E$ is fixed, the collision map operates on a
two-dimensional surface, which, given that the collision takes place on
disk $n$, is conveniently parameterized by the Birkhoff
coordinates $(\phi_n, \xi_n)$, where $\phi_n$ specifies the
angle along the border of disk $n$ that the trajectory makes at collision
with it, and $\xi_n$ is the sinus of the angle that the particle velocity makes
with respect to the outgoing normal to the disk after the collision. 

\subsection{Non-Equilibrium Stationary State}

In order to set up a non-trivial stationary state, we assume that a flux of
trajectories is continuously flowing through the boundaries. This can be
achieved by putting the boundaries in contact with particle
reservoirs  such that the phase-space density at $x=\pm L/2$ is fixed
according to constant values,
\begin{equation}
  \rho(\Gamma, t)\Big|_{x=\pm L/2} = \rho_{\pm},
  \label{lgrhobc}
\end{equation}
corresponding to particle injection rates from the two ends of the channel
at $x=\pm L/2$. All the injected particles have the same energy $E = (v_x^2 +
v_y^2)/2$ and uniform distributions of velocity angles.
The evolution of the phase-space density $\rho$ is otherwise specified by
Liouville's equation, or, equivalently, by the action of the
Frobenius-Perron operator, $\widehat{P}^t$, determined according to
\begin{equation}
  \rho(\Gamma, t) = \widehat{P}^t  \rho(\Gamma, 0)
  \equiv \int_\mathbb{C} d\Gamma'  \delta(\Gamma - \Phi^t\Gamma') \rho(\Gamma', 0)\,.
  \label{pfeq}
\end{equation}

A remarkable result \cite{Gas97} is that the invariant solution of this
equation, compatible with the boundary conditions (\ref{lgrhobc}), is given,
for almost every phase point $\Gamma$, by 
\begin{widetext}
\begin{eqnarray}
  \rho(\Gamma)
  &=& \frac{\rho_++\rho_-}{2} + \frac{\rho_+-\rho_-}{L} 
  \left[x(\Gamma) + \int_0^{-T(\Gamma)} dt\, \dot x(\Phi^t\Gamma)\right],
  \nonumber\\
  &=& \frac{\rho_++\rho_-}{2} 
  \label{lgss}
  + \frac{\rho_+-\rho_-}{L} 
  \left\{x(\Gamma) + \sum_{k=1}^{K(\Gamma)} 
    [x(\Phi^{-t_k}\Gamma) - x(\Phi^{-t_{k-1}}\Gamma)]\right\}.
\end{eqnarray}
In these expressions, $x(\Gamma)$ denotes the projection of the phase point
$\Gamma$ on the horizontal axis, and $T(\Gamma)$ is the time it takes
the phase point $\Gamma$ to reach the system boundary backward in time,
\emph{i.e.} such that $x(\Phi^{-T(\Gamma)}\Gamma) = \pm L/2$. As written in
the second line, the time $T(\Gamma)$ elapses in $K(\Gamma)$ successive
collisions 
separated by time intervals $\tau_n$, such that $t_k$ is the time elapsed
after $k$ collision events, \emph{viz.} $\sum_{j=1}^{k}
\tau_j = t_k$. Obviously $t_{K(\Gamma)} = T(\Gamma)$. The difference
$x(\Phi^{-t_k}\Gamma) - x(\Phi^{-t_{k-1}}\Gamma)$ is the displacement
along the $x$ axis between two successive collisions.

In the limit of infinite number of cells $N$, the number of collisions for
the trajectory to reach the boundaries becomes infinite so that the invariant
density can be written
\begin{eqnarray}
  \rho(\Gamma) &=& 
 \frac{\rho_++\rho_-}{2} 
 \label{lginfss}
 + 
  \frac{\rho_+-\rho_-}{L} 
  \left\{x(\Gamma) + \sum_{k=1}^{\infty} 
    [x(\Phi^{-t_k}\Gamma) - x(\Phi^{-t_{k-1}}\Gamma)]
  \right\}\,.
\end{eqnarray}
\end{widetext}
The two contributions to this expression represent the mean linear density
profile along the axis, plus fluctuations. Notice that the linear profile
is itself the sum of two contributions, the first one being the equilibrium
density and the second one a gradient term. These stationary states are
known after Lebowitz and MacLennan \cite{Leb59,McL59}.
As noted in \cite{Gas97}, the fluctuations form
the singular part of the invariant density, which builds up according to
which of the two boundaries the phase point is mapped to.

\subsection{Continuum Limit}

The linear part of the invariant density profile (\ref{lgss}) has its
origin in the 
diffusion process which takes place at the phenomenological level. Indeed,
in the continuum limit, the phase-space density reduces to the macroscopic
density distribution $\mathcal{P}(X,t)$, whose evolution is described by
Eq.~(\ref{FPeq}), with macroscopic position variable $X$ and diffusion
coefficient, $\mathcal{D}$, which can, in principle, be determined from
the underlying dynamics. 

In order to obtain Eq.~(\ref{FPeq}) from the Liouvillian evolution,
Eq.~(\ref{pfeq}), we let $l\to0$ and $N\to\infty$ with
$Nl = L$ fixed. All other quantities are fixed, in particular the particle
velocity, which is taken to be unity in the length units of $L$. In that
limit, the macroscopic density can be obtained from the phase-space density
by averaging over phase-space regions such that the position variables
$x(\Gamma)$ are identified with the macroscopic position variable $X$, here
denoted $X_n$ and identified with the cell $\mathbb{C}_n$,
\begin{equation}
  \mathcal{P}(X_n, t) = 
  \frac{1}{l}\int_{\mathbb{C}_n}d\Gamma
  \rho(\Gamma,t)\,.
  \label{ccP}
\end{equation}
Note that the volume element is here and throughout assumed to be properly
normalized, \emph{i.e.} $\int_{\mathbb{C}_n}d\Gamma = l$.

Given non-equilibrium flux boundary conditions with the stationary state
(\ref{lginfss}),  the only contribution to the particle density 
$\mathcal{P}(X_n)$ is
the linear part, which, in the continuum limit, yields the stationary
solution to Eq.~(\ref{FPeq}), namely  
\begin{equation}
  \mathcal{P}(X) = \frac{\mathcal{P}_++\mathcal{P}_-}{2}
  + \frac{X}{L}(\mathcal{P}_+-\mathcal{P}_-)\,,
  \label{FPeqss}
\end{equation}
where $\mathcal{P}_\pm$ is equal to $\rho_\pm$ up to a volume factor,
and specifies the boundary conditions $\mathcal{P}(\pm L/2)$.

The singular part of the invariant density, on the other hand, has no
macroscopic counterpart. Its integration over the cell indeed vanishes by
isotropy of the underlying dynamics. Nonetheless, as we will demonstrate
shortly, it bears essential information pertaining to the microscopic
origin of entropy production. 

In that respect, the phenomenological entropy production rate (\ref{thermoep})
can be written
\begin{eqnarray}
  \frac{d_\mathrm{i}\mathcal{S}(X)}{dt} &=& \mathcal{D}
  \frac{\left[\partial_X \mathcal{P}(X)\right]^2}{\mathcal{P}(X)},
  \label{epolg}\\
  &=& \frac{\mathcal{D}}{L^2}\frac{(\mathcal{P}_+ - \mathcal{P}_-)^2}
  {\mathcal{P}_\mathrm{eq}
    + \frac{X}{L}(\mathcal{P}_+-\mathcal{P}_-)}\, ,
  \nonumber
\end{eqnarray}
where we denoted $\mathcal{P}_\mathrm{eq} =
(\mathcal{P}_++\mathcal{P}_-)/2$.

\subsection{\label{sec.lgnr}Numerical Results}

We can verify numerically that the invariant density of the Lorentz channel
has the form (\ref{lgss}), with a linear part given by the
stationary solution  of the Fokker-Planck equation (\ref{FPeqss}), and a
singular fluctuating part. For the sake of the numerical computation, we
let $L \equiv 1$ and consider a channel with $2N+1$ disks, $N=25$, the
left- and right-most ones being only half disks. The macroscopic position
$X$, $-L/2\leq X\leq L/2$, will be identified with the cells $\mathbb{C}_n$
about the corresponding disks, letting $X_n = n l/2$, $-N\leq n \leq N$. We
then fix the injection rates $\rho_\pm$ and let $\rho_- >0$ and $\rho_+ =
0$ so as to have $\mathcal{P}_- = 1$ and $\mathcal{P}_+ =0$ for the
corresponding macroscopic quantities.

Having fixed the geometry of the channel and the injection rates, we
compute the invariant phase-space density $\rho(\Gamma)$ in terms of the
corresponding density of the collision map, also called the Birkhoff map,
which takes trajectories from one collision event with a disk to the next
one. This is indeed much easier 
since the numerical integration of the Lorentz gas relies on an event
driven algorithm corresponding to integrating the collision map. The
conversion between the two phase-space densities is here trivial because
the time scales are uniform. 

\begin{figure}[thb]
  \centering
  \includegraphics[angle=0,width=.45\textwidth]{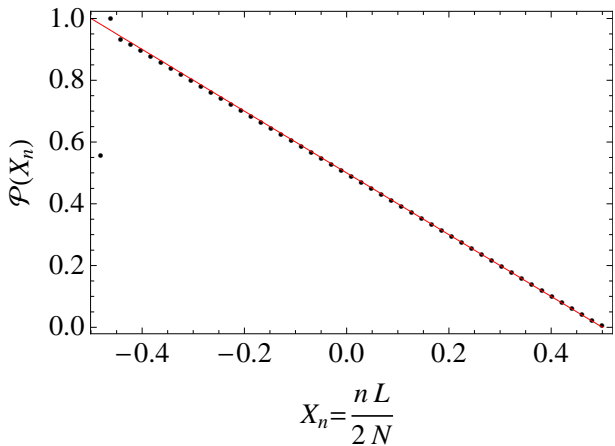}
  \caption{Non-equilibrium stationary density of the Lorentz channel
    obtained for a cylinder of length $L = 1$, with $51$ disks ($N =
    25$). The 
    solid line is $\mathcal{P}(X) = 1/2 - X$, the solution of
    Eq.~(\ref{FPeqss}), with $\mathcal{P}_- = 1$ and $\mathcal{P}_+ =
    0$. Notice that the left- and right-most disks are only half
    disks. Hence the strong boundary effects.}
  \label{fig.olgss}
\end{figure}

Thus the linear part of the invariant density is computed by considering a
large set of initial conditions injected into the channel from the left end
and integrating them until they reach either ends of the channel, thus
exiting the system. In the meantime, we record the number of collisions
each particle performs with each disk, the ensemble average of which
approximates the stationary distribution $\mathcal{P}(X_n)$.
Provided $N$ is sufficiently large and $n$ is not too close to the channel
boundaries, this ensemble average is expected to approximate
$\mathcal{P}(X)$, as specified by 
Eq. (\ref{FPeqss}). The result is displayed in Fig. \ref{fig.olgss} and was
adjusted by an overall constant 
so as to match $\mathcal{P}_- = 1$ (and $\mathcal{P}_+ = 0$). The agreement
is very good.

The singular part of the density can also be computed in terms of
the statistics 
of the collision map. For the sake of characterizing this quantity, we
notice that the equilibrium Lorentz gas (a single disk on an hexagonal cell
with periodic boundary conditions) preserves the Liouville measure 
\begin{eqnarray}
  d\Gamma &=& dx\, dy\,  dv_x\,  dv_y\,,\nonumber\\
  &=& dE\,  dt\,  d\phi\, d\xi,
  \label{lgbirk}
\end{eqnarray}
where $\phi$ and $\xi$ are the Birkhoff coordinates.

\begin{figure*}[tbph]
  \centering
  \includegraphics[angle=0,width=.4\textwidth]{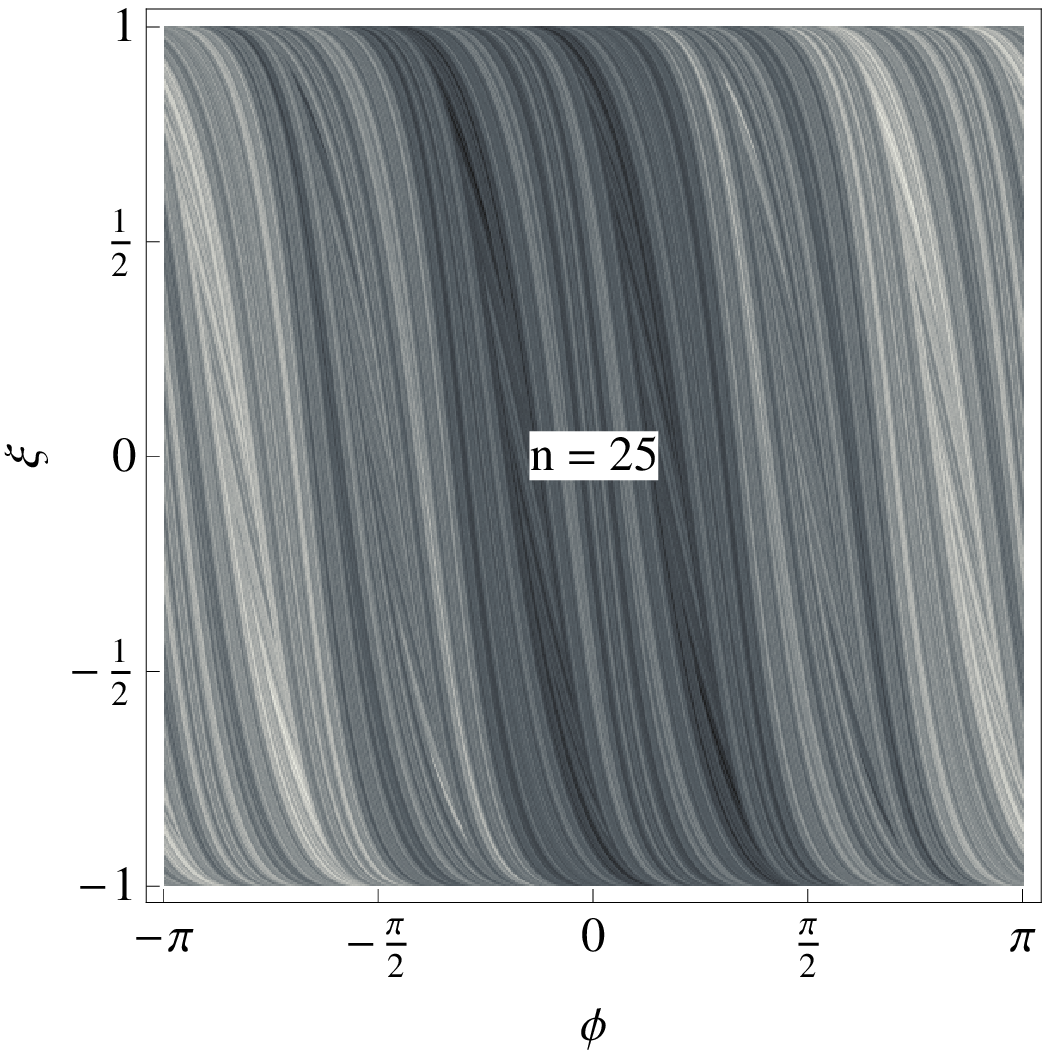}
  \hfill
  \includegraphics[angle=0,width=.4\textwidth]{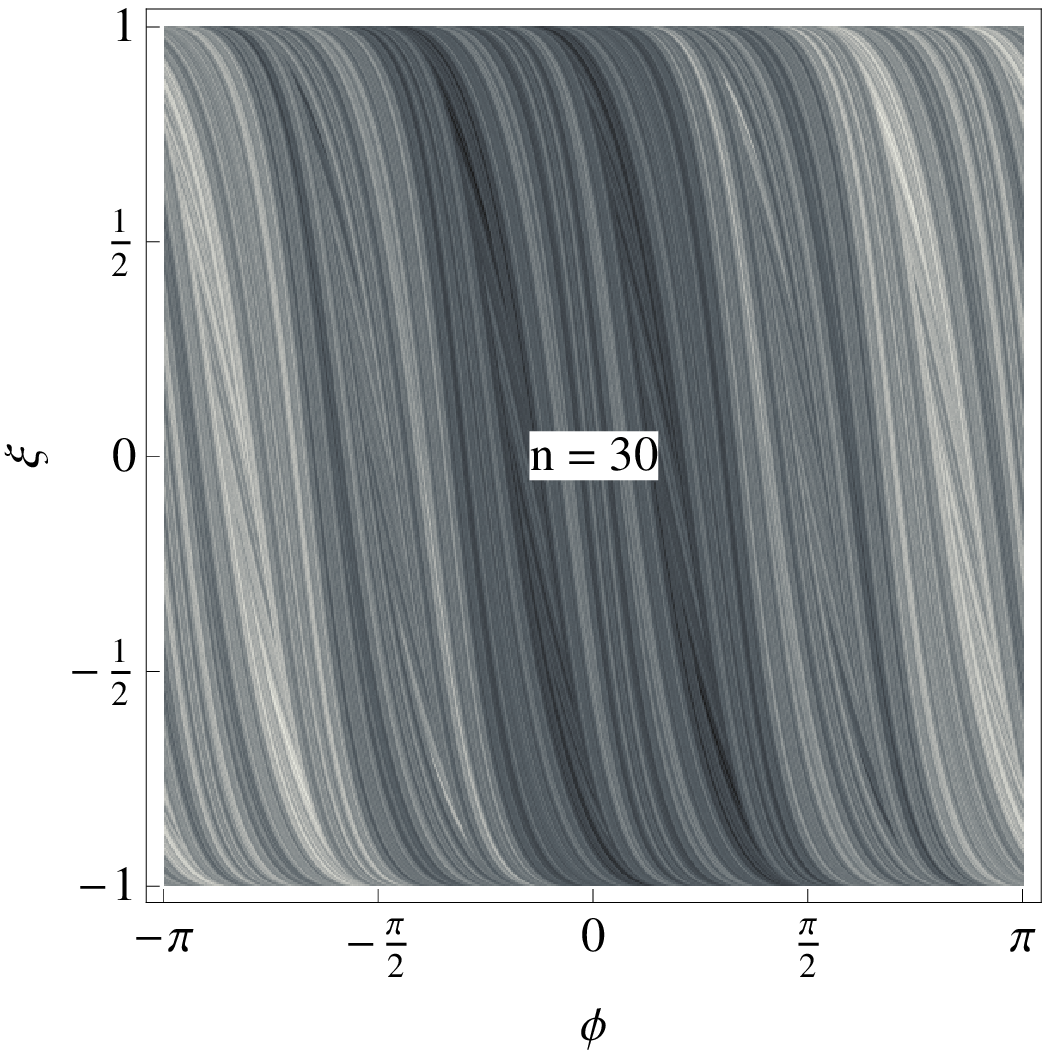}
  \includegraphics[angle=0,width=.4\textwidth]{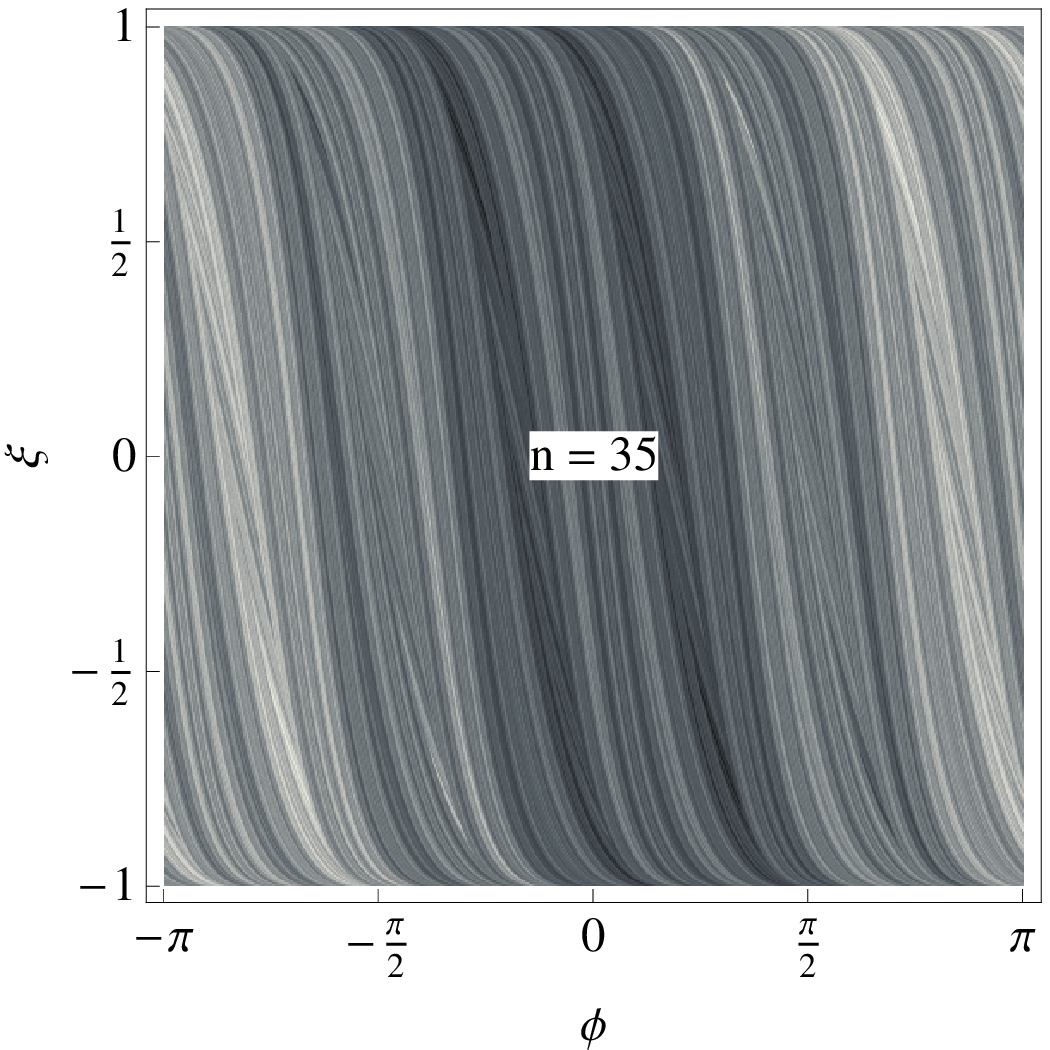}
  \hfill
  \includegraphics[angle=0,width=.4\textwidth]{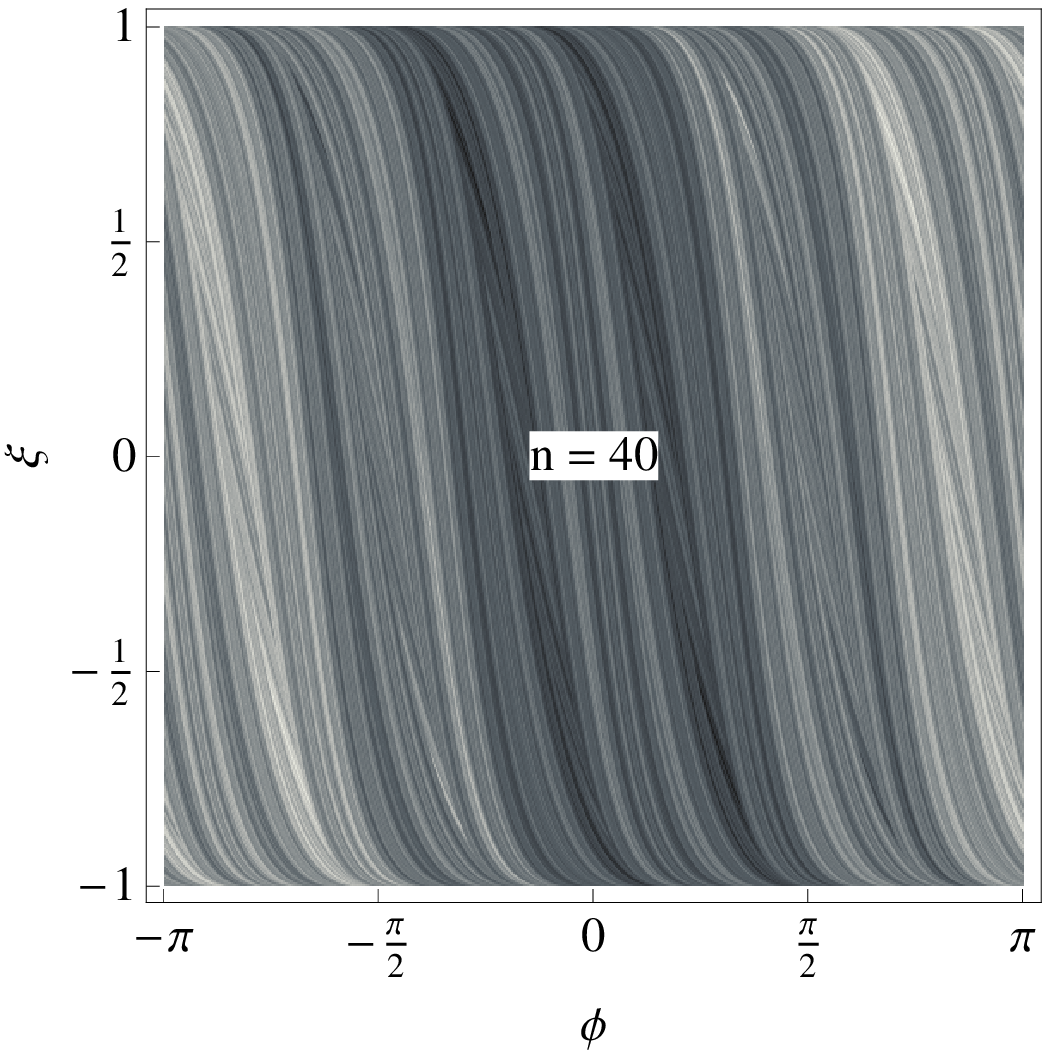}
  \includegraphics[angle=0,width=.4\textwidth]{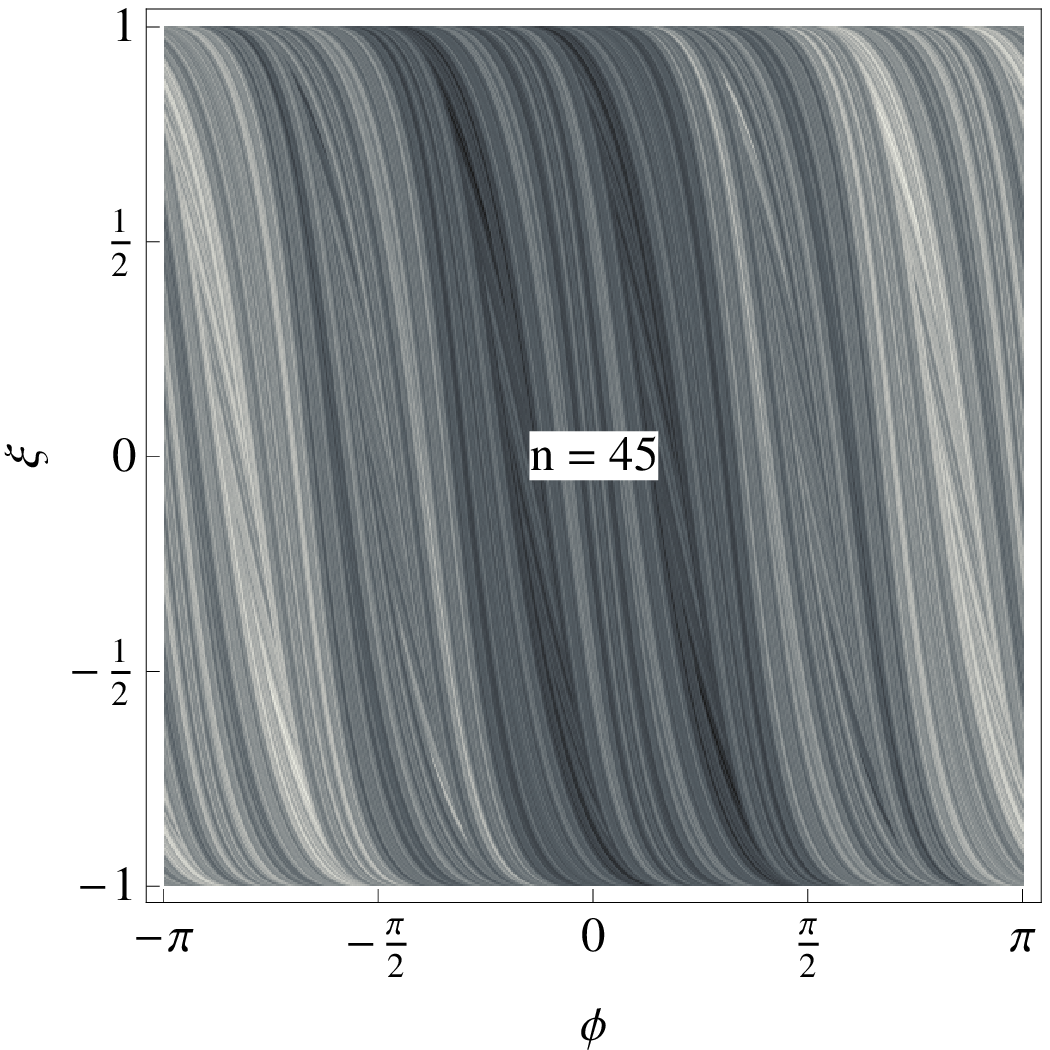}
  \hfill
  \includegraphics[angle=0,width=.4\textwidth]{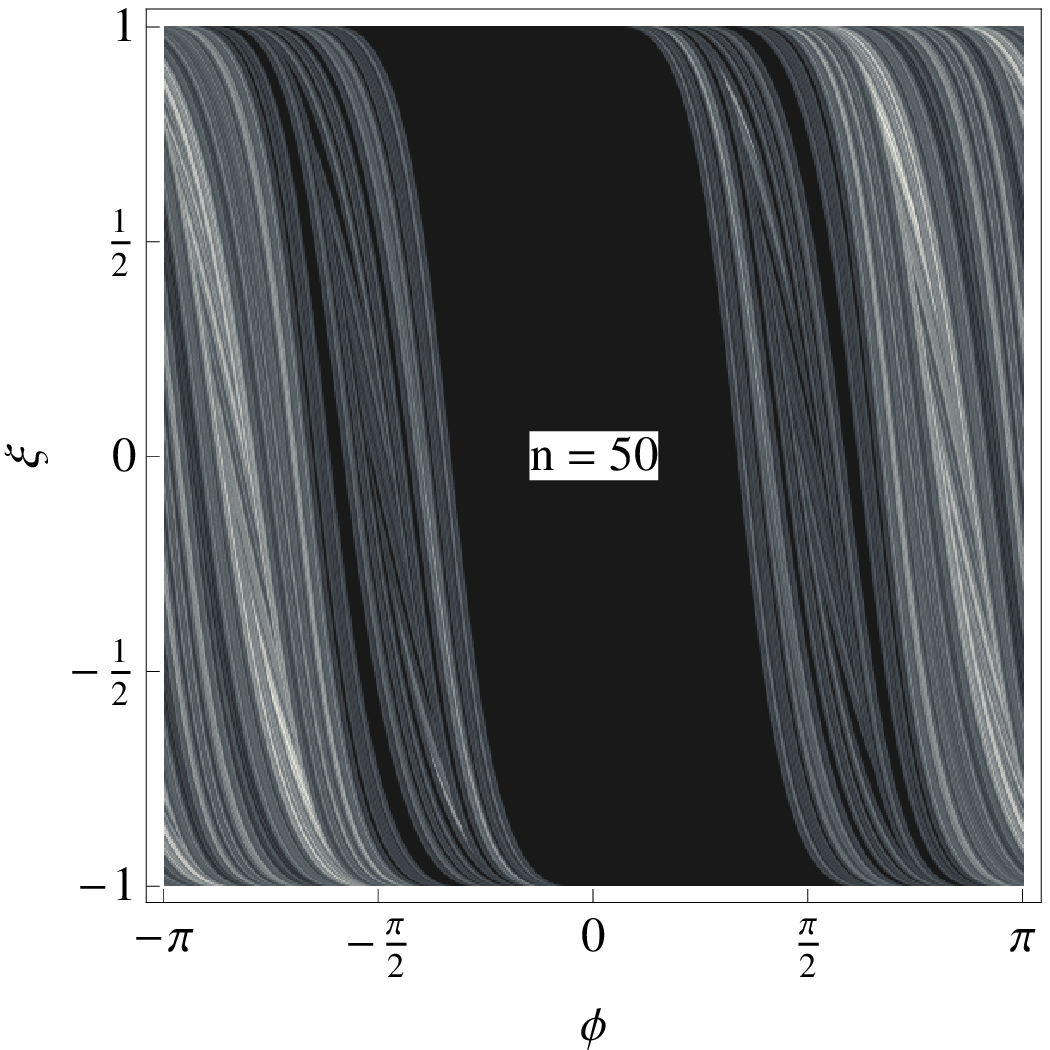}
  \caption{Fractal structure of the phase portraits of an open cylindrical
    Lorentz channel with absorbing boundaries at its ends. The
    plots are histograms computed over grids of $500\times 500$ cells,
    and counting the average collision rates of many trajectories in every
    cell. Disk $50$ is the one before last. The geometry of the
    channel is that shown Fig.~\ref{fig.lg}. Here
    particles are injected at the left boundary only. The color white,
    associated to higher counting rates, corresponds to 
    the injection of particles from the left boundary.  
    Black areas, on the other hand, correspond to absorption at the right
    boundary. Hues of gray are associated to phase-space
    regions whose points are mapped backward to both left and right
    borders.  The corresponding overall densities (\ref{ccP}) are shown in 
    Fig.~\ref{fig.olgss}.
}  
  \label{fig.olgpp}
\end{figure*}

For the open Lorentz gas, we associate to each disk a pair of Birkhoff
coordinates, $(\phi_n,\xi_n)$, and record each collision event in a
histogram. The results, displayed in Fig.~\ref{fig.olgpp}, show the
fractality of the fluctuating part of the invariant density, with boundary
effects vanishing exponentially fast with respect to the distance of the
disk to the channel boundaries.

We will see in Sec. \ref{sec.ep} that the fractality of the stationary
measure displayed in Fig.~\ref{fig.olgpp} is responsible for the entropy
production rate that is associated to the mass current in the open Lorentz
gas with flux boundary conditions.

\section{\label{sec.mbm}Multi-baker map}

The analytic derivation of the entropy production rate (\ref{epolg}),
relying on the fractality of the invariant measure (\ref{lginfss}), can be
achieved with the multi-baker map. This is a simple model of deterministic
diffusion that can be thought of as a caricature of the dynamics of the
Lorentz gas described above. This model was originally introduced in
\cite{Gas92}, its non-equilibrium stationary state and relation to
thermodynamics analyzed in \cite{TG95}, and the derivation of the entropy
production rate described in \cite{Gas97a}.

\begin{figure}[thb]
  \centering
  \includegraphics[angle=0,width=.45\textwidth]{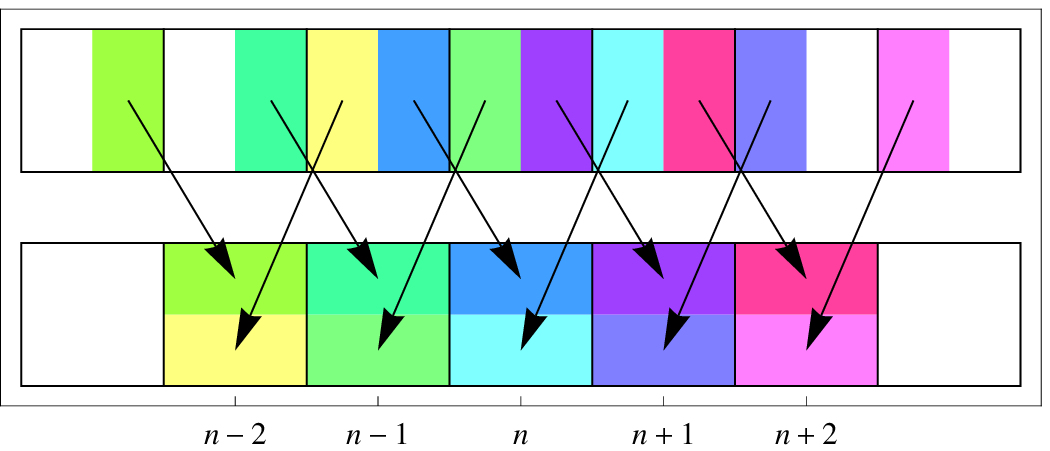}
  \caption{(Color online) Multi-baker map (\ref{mbmap}). Each cell has area $l^2$, with
    coordinates $(x,y)$, and is labeled by an integer $-N/2\leq n \leq N/2$.}
  \label{fig.mbmap}
\end{figure}

The basic idea which underlies the similarity with the Lorentz channel is
threefold. First, let the positions take discrete lattice positions, like
the cells $\mathbb{C}_n$ of the channel;
second, assume that the collision times $\tau_n$ are all identical and
denoted by $\tau$, which we leave arbitrary for now; and, third, replace the
reflection rules at the collisions by a simple Bernoulli-type
angle-doubling rule, which decides which direction the particle
goes to. Assuming the system length $L$ to be an integer multiple of the
cell size $l$, which for the sake of the argument we take to be $L =
(N+1)l$ with $N$ even, tracer dynamics of the Lorentz channel are
replaced by the mapping on $\{-N/2,\dots,N/2\}\otimes[0,l]^2$, 
given by 
\begin{eqnarray}
  \lefteqn{B_0:\,(n,x,y) \mapsto }
  \label{mbmap}\\
  &&\left\{
    \begin{array}{l@{\quad}l}
      \big(n - 1, 2x, y/2 \big)\,,& 0\leq x < l/2\,,\\
      \big(n + 1, 2x - l, (y + l)/2 \big)\,,& l/2\leq x < l\,,
    \end{array}
  \right.
  \nonumber
\end{eqnarray}
with the convention that trajectories end when mapped to $n$ outside of
$\{-N/2,\dots,N/2\}$. The map is displayed in Fig.~\ref{fig.mbmap}. 
We point out that the multi-baker map has a time-reversal symmetry with
respect to the operator $S:(n,x,y)\mapsto (n,l-y,l-x)$, namely $S\circ B_0
= B_0^{-1}\circ S$. The map (\ref{mbmap}) includes a displacement from cell
to cell, which we denote by $\Lambda$, 
\begin{equation}
  \Lambda(n,x,y) \equiv
  \left\{
    \begin{array}{l@{\quad}l}
      - l\,,& 0\leq x < l/2\,,\\
      + l\,,& l/2\leq x < l\,.
    \end{array}
  \right.
  \label{mbdisp}
\end{equation}

Because $B_0$ is a Bernoulli map, a point $\Gamma = (n,x,y)$
in phase-space may be thought of as coding an infinite sequence of
equi-probable nearest-neighbors jumps both in the past and future, with
initial condition at position $nl$. Moreover the variables $(x,y)$, play
here the role of the Birkhoff coordinates $(\phi,\xi)$ of the Lorentz gas.

\subsection{Non-Equilibrium Stationary State}

Given flux boundary conditions, a non-equilibrium stationary state similar
to that of Eq.~(\ref{lgss}) sets in, with a singular part given in terms of
the sum of the successive displacements the tracer makes before it reaches the
outer boundaries and gets absorbed. The $x$ position now
becomes a lattice coordinate $n$ and the change in positions between
collisions the displacement (\ref{mbdisp}), so that the stationary state 
density reads
\begin{eqnarray}
  \lefteqn{\rho(\Gamma)}\label{mbss}\\
  &=& \frac{\rho_++\rho_-}{2} + \frac{\rho_+-\rho_-}{L} 
  \left[n(\Gamma)l + \sum_{k=1}^{K(\Gamma)} \Lambda (B_0^{-k}\Gamma)\right],
  \nonumber
\end{eqnarray}
where $n(\Gamma)$ denotes the projection of $\Gamma$ along the integer axis.

Since the sum over the successive displacements is a highly singular
function of $\Gamma$, it is convenient to introduce a better behaved
characterization of the invariant density in terms of cumulative
functions. Thus let $\Gamma = (n, x, y)$ with $0\leq x,y < l$.
We define the set\footnote{Such sets are usually referred to in the
  literature as cylinder sets.}
\begin{equation}
  \mathbb{C}_n(x, y) = \{ (x', y') \,|\, 
  0\leq x' < x,\,0\leq y'<y\} .
\end{equation}
For $x=y=l$, the cylinder set corresponds to the whole cell,
$\mathbb{C}_n\equiv\mathbb{C}_n(l,l)$. The cumulative measure of this set
is defined as 
\begin{eqnarray}
  \mu_n(x,y) &=& 
  \int_{\mathbb{C}_n(x,y)} d\Gamma \rho(\Gamma)\,,\nonumber\\
  &=& \left[\frac{\rho_++\rho_-}{2} + (\rho_+-\rho_-)\frac{nl}{L}\right]
  \frac{x y}{l^2}
  \label{mbmapmunxy}\\
  &&+\frac{\rho_+-\rho_-}{L} \int_{\mathbb{C}_n(x,y)}d\Gamma
  \sum_{k=1}^{K(\Gamma)} \Lambda (B_0^{-k}\Gamma)\,.
  \nonumber
\end{eqnarray}
Here we again assumed normalization of the volume element, $d\Gamma = 
dx\,dy/l^2$. 

It can be shown \cite{TG95} that the expression above reduces to
\begin{equation}
  \mu_n(x,y)  = \mu_n \frac{x y}{l^2} 
  +\frac{(\rho_+-\rho_-)}{L}x T_n\left(\frac{y}{l}\right)\,,
\end{equation}
where  the linear part of the invariant density is
\begin{equation}
\mu_n \equiv \mu_n(l,l) = \frac{1}{2}(\rho_++\rho_-) + 
(\rho_+-\rho_-)\frac{nl}{L}\,,
\label{mbmapmun}
\end{equation}
and $T_n$ is the incomplete Takagi function \cite{TG95}, which can be
defined through the functional equation, here with $y \in [0, 1]$, 
\begin{equation}
  T_n(y) = \left\{
    \begin{array}{l@{\quad}l}
      y + \frac{1}{2}T_{n+1}(2 y)\,,& 0\leq y < 1/2,\\
      1-y + \frac{1}{2}T_{n-1}(2 y - 1)\,,& 1/2\leq y < 1.
    \end{array}
  \right.
  \label{inctakagi}
\end{equation}
This equation can be solved recursively using dyadic expansions of $x$ and
the boundary conditions $T_{\pm N/2}(x) = 0$. Notice that the size $l$ of
the jumps $\Lambda$ was extracted from the Takagi functions and absorbed
into $x$ ($\equiv l\times x/l$)\,.

We remark that $\mu_n$ verifies 
\begin{equation}
  \mu_n = \frac{1}{2}\mu_{n-1} + \frac{1}{2}\mu_{n+1},
  \label{rwsym}
\end{equation}
which describes the invariant statistics of a uniform random walk with jump
probabilities $1/2$ to nearest neighbors. Given the length $l$ of the
jumps, and their rate $\tau$, the diffusion coefficient for this process is
$\mathcal{D} = l^2/(2\tau)$. The proper diffusive scaling is recovered
provided $\tau \sim l^2$ when $\tau, l \to 0$ in the continuum limit.

\begin{figure}[thb]
  \centering
  \includegraphics[angle=0,width=.45\textwidth]{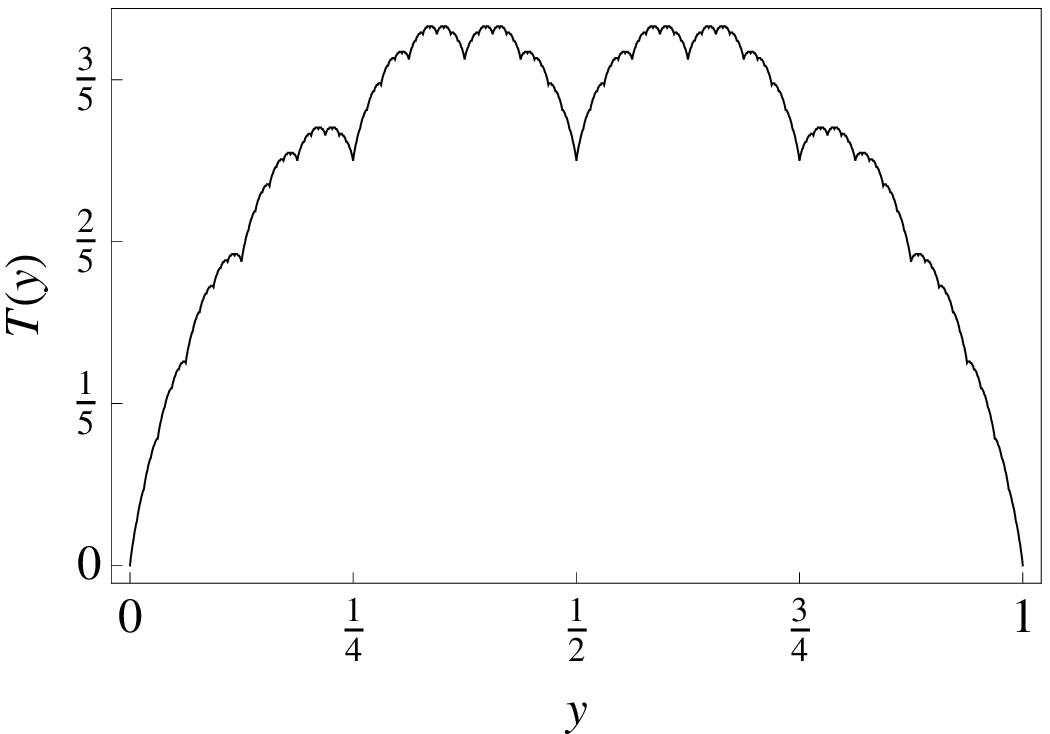}
  \includegraphics[angle=0,width=.45\textwidth]{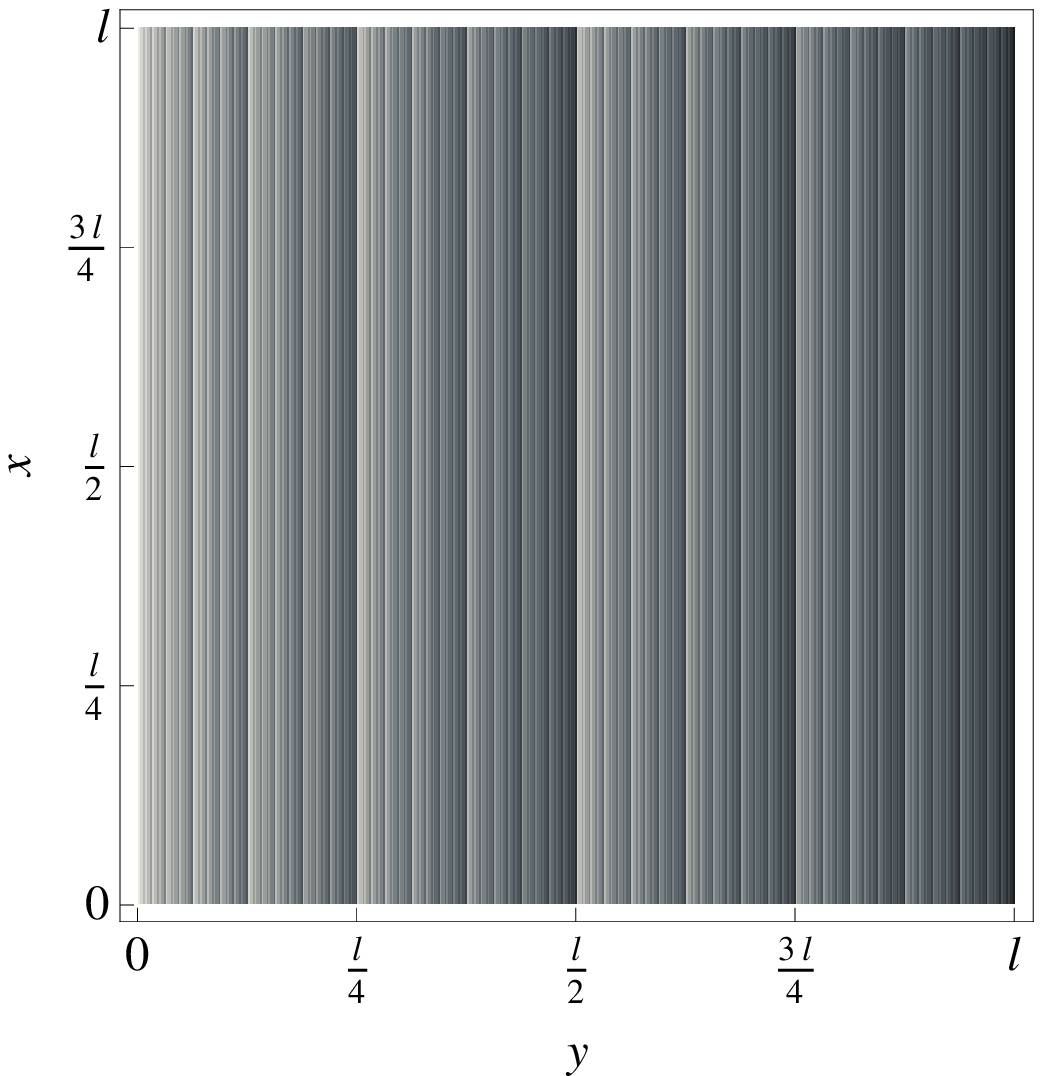}
  \caption{(Top) The Takagi function, Eq.~(\ref{takagi}), characterizes the
    fluctuations of the non-equilibrium stationary state associated to the
    multi-baker map under flux boundary conditions. (Bottom) The
    singularity of the invariant density of the multi-baker map
    (\ref{mbss}) can be visualized by measuring the local slopes of the
    Takagi function (\ref{takagi}) at a given scale, here $dy =
    1/512$. This histogram can be compared to those obtained for the
    Lorentz channel, Fig.~\ref{fig.olgpp}.}
  \label{fig.takagi}
\end{figure}

In the continuum limit, the number of steps that separate phase points from
the boundaries becomes infinite so that the invariant cumulative measure
can be written in terms of the (complete) Takagi function,
\begin{equation}
  \mu_n(x,y)  = \mu_n \frac{x y}{l^2} +
  \frac{(\rho_+-\rho_-)}{L}\,x\, T\left(\frac{y}{l}\right)\,,
  \quad(N\to\infty)\,.
  \label{mbmapmuninf}
\end{equation}
The Takagi function \cite{Tak1903}, displayed on the top panel of
Fig.~\ref{fig.takagi}, is the solution of the functional equation 
\begin{equation}
  T(y) = \left\{
    \begin{array}{l@{\quad}l}
      y + \frac{1}{2}T(2 y)\,,& 0\leq y < 1/2,\\
      1-y + \frac{1}{2}T(2 y - 1)\,,& 1/2\leq y < 1.
    \end{array}
  \right.
  \label{takagi}
\end{equation}
This equation, due to de Rham \cite{Rham57}, is one of many equivalent
representations of the Takagi function, $T(y) = \sum_{n=0}^\infty
2^{-n}\,|2^n y - [2^n y + 1/2]|$, where $[x]$ stands for the maximum
integer not exceeding $x$. It is an everywhere continuous function, but has
no bounded derivative.

The singularity of the fluctuating part of the invariant density
(\ref{mbss}) with respect to the $y$ component can be visualized by
measuring the local slopes of the Takagi function. The dependence on the $x$
component on the other hand is trivial. The resulting histogram, displayed
on the bottom panel of Fig.~\ref{fig.takagi}, can be compared to those
shown in Fig.~\ref{fig.olgpp} for the Lorentz channel. The main difference
is that while the fractal structures are curved in the open Lorentz system,
they are parallel to the $x$-direction in the multi-baker map. The reason
is that the fractal structures are smooth with respect to the unstable
directions, which, for the multi-baker map, are always directed along the
$x$-axis, while their direction varies from point to point in the Lorentz
system. 

\subsection{Continuum Limit}

As with the Lorentz gas, we let $l\to0$ and $N\to\infty$, in the continuum
limit, keeping $L = (N+1)l$ constant. In this limit, the macroscopic 
particle density is to be identified with $\mu_n$, $\mathcal{P}(X_n = nl)
= \mu_n$. On the scale of $\mu_n$, the singular fluctuations embodied by the
Takagi function disappear. However, as argued in \cite{Gas97a}, these
fluctuations are responsible for the positiveness of the entropy production
rate, given according to phenomenology by Eq.~(\ref{epolg}). We will
comeback to this identification in Sec.~\ref{sec.ep}.

\section{\label{sec.sps}Spatially periodic systems of interacting
  particles}  

The results presented in Secs. \ref{sec.olg}-\ref{sec.mbm} for diffusive
systems of non-interacting tracer particles can be extended to spatially
periodic systems of many interacting particles where tagged particles
--assumed to be independent of each other-- undergo a process of
self-diffusion. The arguments below provide an extension to non-equilibrium
stationary states of the results presented in \cite{DGG02} for tagged
particle diffusion in interacting particle systems undergoing a relaxation
to equilibrium. 

We follow the diffusion of tracer particles in a periodic array of cells,
each containing $Q$ particles, all of them moving with interactions. To
this purpose we make the assumption that the density of tracer particles is
\emph{dilute}, so that we can consider each tracer particle as though it
were the only such particle in each cell at any given time. This is
consistent with processes of self-diffusion and mutual diffusion when the
tracer is not identical to the other particles. The issue of considering
many tagged particles with correlated motions, such as in a process of
color diffusion will be discussed in the Conclusions. 

Each cell of our system has a finite size domain which is
delimited by periodic boundary 
conditions. This domain can for instance be a square in two dimensions or a
cube in three dimensions. The center of mass is taken to be at rest so that
no net current takes place. For the sake of the argument, we will assume
that the periodic array of cells has a cylindrical structure, extended
along one of its spatial dimensions and periodic otherwise. 

The dynamics is thus constructed in a way that it is periodic from one
cell to the other. A particle entering one cell through any of its borders
has a corresponding image which simultaneously leaves that cell through the
opposite border, entering the corresponding neighboring cell. The cells
located at the boundaries of the array have however 
a special status since some of their borders are not contiguous to a
neighboring cell. Particles can therefore leave the system through these
borders, and thus be absorbed, while other particles are correspondingly
being injected through the opposite side of the cylinder, entering the
corresponding cell through the opposite border. While this process occurs
at constant rate, it can nevertheless be used to induce a mass current by
simply tagging one of the particles at a time with prescribed rates
of injection into the system. This way a molecular dynamics simulation of
tracer particle diffusion in a spatially extended system can be set up,
which involves a finite number of degrees of freedom only. 

Thus consider the cubic box of side $l$, $\mathbb{I} = [-l/2,l/2]^d \subset
\mathbb{R}^d$, containing a collection of $Q$ non-overlapping hard sphere
particles of radii $\sigma$ with phase-space coordinate $\Gamma_Q =
\{(\vec{q}_i, \vec{p}_i)\}_{1\leq i \leq Q}$, $\vec{q}_i \in \mathbb{I}$,
$\vec{p}_i \in \mathbb{R}^d$. Let $\mathbb{C}_n$ denote the phase space of the
$n$th elementary cell. The local dynamics, denoted $\phi^t\Gamma_Q$,
preserves the micro-canonical measure and is assumed to be chaotic and
mixing. 

\begin{figure}
  \centering
  \includegraphics[angle=0,width=.45\textwidth]{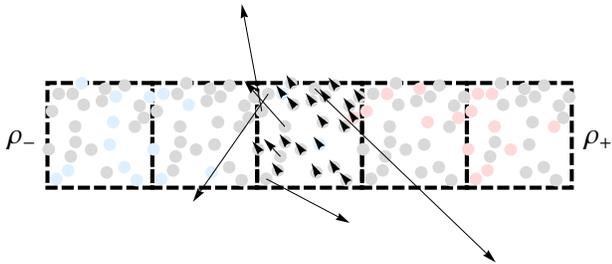}
  \caption{Example of a spatially periodic system with many interacting
    particles and one tagged particle in each cell. The tagged particles are
    randomly injected at the left and right boundaries with suitable
    probabilities $\rho_\pm$. For purposes of visualization, the number of
    tracer particles has been exaggerated, mimicking a dilute system where
    different tracers move independently of one another.
  }
  \label{fig.sps}
\end{figure}
Given a collection of $N$ copies of such a box, which are placed side by
side along the first coordinate axis, the position of a given tracer
particle is specified by its coordinate $\vec{q} = (q^1, \dots, q^d)$
within the elementary box, identical to one out of the $Q$ particles there,
and by its lattice position $n$, $-N/2 \leq n \leq N/2$ (we assume $N$
even), which distinguishes it from its images in the other lattice
cells. Its real coordinate is therefore $(q^1 + n l, \dots, q^d)$. The
flow $\Phi^t\{n, \Gamma_Q\}$ acts over the extended system
$\cup_n\mathbb{C}_n$, 
\emph{i.e.} on the $NQ$ particles of the $N$ elementary cells, out of which
a single particle --say the first among $Q$-- has been tagged, the
corresponding image being located in cell $n$, until it moves to a
neighboring cell under the time evolution $\Phi^t$ . Figure \ref{fig.sps}
shows an example of such a system where the number of tracer particles has
however been exaggerated 
for visual purposes. We are indeed assuming a dilute limit according to
which no correlations take place between tracer particles.

\begin{widetext}
Given tracer injection rates $\rho_-$ and $\rho_+$ at the left and right
borders, the stationary tracer density is written in a way similar to
Eq.~(\ref{lginfss}), in the form
\begin{equation}
  \rho(\{n, \Gamma_Q\})
  = \frac{\rho_++\rho_-}{2} 
 + \frac{\rho_+-\rho_-}{L} 
  \left\{L(\{n,\Gamma_Q\}) + \sum_{k=1}^{\infty} 
    \Big[L(\Phi^{-t_k}\{n,\Gamma_Q\}) - L(\Phi^{-t_{k-1}}\{n,\Gamma_Q\})
    \Big]
  \right\}\,.
  \label{spsinfss}
\end{equation}
\end{widetext}
This density is assumed to be normalized with respect to the local
micro-canonical distribution of $Q$ particles per cell, and the tracer
position along the cylinder axis is $L(\{n,\Gamma_Q\}) = nl+ q^1$. We can
take the discrete position $L$ instead of the continuous 
position along the horizontal axis because both give equivalent results
since the fractal structures appear through the continuous phase-space
dependence of $\rho$ on $\Gamma_Q$. The factorization occurs because the
tracer dynamics is a simple 
passive advection process under the assumption that tracer particles do not
interact. Thus equilibration of the interacting 
particles takes place regardless of the tracer dynamics.

The tracer density (\ref{spsinfss}) is a conditional distribution of
finding a single tagged particle at the phase point $\Gamma_Q$ in cell $n$
given the relative injection rates at the system's boundaries. This density
is defined with respect to the micro-canonical equilibrium distribution in
every cell. The continuum limit is recovered, as with the 
open Lorentz gas and multi-baker map, by considering an infinite number of
copies of independent systems, each with a single 
tagged particle. The average number of tagged particles as a function of
the position along the position coordinate thus yields a macroscopic
density which evolves according to the Fokker-Planck equation
(\ref{FPeq}). The tracer dynamics we consider here is therefore much like a
single particle system, except for the dimensionality of the phase-space
dynamics. In particular the diffusion coefficient depends on the collective
motion of the $Q$ interacting particles per periodic cell. 

A more physically consistent limit would be to consider the absence of
correlations between tracer 
particles set in motion in the same arbitrary large system. However our
formalism for the identification of fractal structures in the
non-equilibrium stationary states and the computation of the entropy
production rate makes systematic use of the strict spatial periodicity of 
the dynamics. Its extension to such more general situation requires
appropriate elaborations. We leave it as a perspective for the time being.

\section{\label{sec.ep}Entropy Production}

Though a fine-grained Gibbs-type entropy associated to a time-dependent
phase-space density $\rho(\Gamma,t)$, 
\begin{equation}
  S^t(\mathbb{C}_n) = -\int_{\Gamma\in\mathbb{C}_n} 
  d\Gamma \rho(\Gamma,t) [\log\,\rho(\Gamma,t) - 1], 
  \label{Gibbsent}
\end{equation}
is preserved by the time evolution, this is not the case for coarse-grained
entropies. This observation has long been understood, see
\emph{e.g.} \cite{Tol39, Ehr59}, however the novelty here is that, even
though the system under consideration is Hamiltonian, the stationary state
distribution has a singular density. The fractal structure of the
stationary state forbids the use of an entropy like (\ref{Gibbsent}).
Accordingly, the proper treatment of these phase-space measures requires
the use of coarse graining methods.

A systematic approach to defining the proper coarse-grained entropy was
outlined in \cite{Gas97a, GD99}. The idea is that, in the stationary state,
owing to the singularity of the invariant density, the Gibbs entropy
(\ref{Gibbsent}) should be evaluated by using a grid of phase space, or
partition, $\mathbb{G} = \{d\Gamma_j\}$, into small volume elements
$d\Gamma_j$, and a time-dependent state $\mu_n(d\Gamma_j, t)$. The entropy
associated to cell $\mathbb{C}_n$, coarse grained with respect that grid, is
defined according to 
\begin{equation}
  S_{\mathbb{G}}^t(\mathbb{C}_n) = - \sum_j
  \mu_n(d\Gamma_j, t) \left[\log\frac{\mu_n(d\Gamma_j, t)}{d\Gamma_j} -
    1\right]. 
  \label{Gibbsentcc}
\end{equation}
This entropy changes in a time interval $\tau$ according to
\begin{eqnarray}
  \Delta^\tau S^t(\mathbb{C}_n) 
  &=& S_{\mathbb{G}}^t(\mathbb{C}_n) - S_{\mathbb{G}}^{t-\tau}(\mathbb{C}_n)
  \,,\label{deltaStau}\\
  &=& S_{\{d\Gamma_j\}}^t(\mathbb{C}_n) - 
  S_{\{\Phi^\tau d\Gamma_j\}}^{t}(\Phi^\tau \mathbb{C}_n)
  \,,
  \nonumber
\end{eqnarray}
where, in the second line, the collection of partition elements
$\{d\Gamma_j\}$ was mapped to $\{\Phi^\tau d\Gamma_j\}$, which forms a
partition $\Phi^\tau\mathbb{G}$ whose elements are typically stretched
along the unstable foliations and folded along the stable foliations. 

Following \cite{DGG02}, and in a way analogous to the phenomenological
approach to entropy production \cite{deGMaz}, the rate of entropy change
can be further decomposed into entropy flux and production terms according
to
\begin{equation}
  \Delta^\tau S_{\mathbb{G}}^t(\mathbb{C}_n) = 
  \Delta_\mathrm{e}^\tau S_{\mathbb{G}}^t(\mathbb{C}_n) 
  + \Delta_\mathrm{i}^\tau S_{\mathbb{G}}^t(\mathbb{C}_n)\,,
  \label{entdecomp}
\end{equation}
where the entropy flux is defined as the difference between the entropy
that enters cell $\mathbb{C}_n$ and the entropy that exits that cell,
\begin{equation}
  \Delta_\mathrm{e}^\tau S_{\mathbb{G}}^t(\mathbb{C}_n) = 
  S_{\{\Phi^\tau d\Gamma_j\}}^t(\mathbb{C}_n) 
  - S_{\{\Phi^\tau d\Gamma_j\}}^t(\Phi^\tau \mathbb{C}_n) \,.
  \label{entchange}
\end{equation}
Collecting Eqs.~(\ref{deltaStau})-(\ref{entchange}), the entropy production
rate at $\mathbb{C}_n$ measured with respect to the partition $\mathbb{G}$,
is identified as
\begin{equation}
  \Delta_\mathrm{i}^\tau S_{\mathbb{G}}^t(\mathbb{C}_n) = 
  S_{\{d\Gamma_j\}}^t(\mathbb{C}_n) 
  - S_{\{\Phi^\tau d\Gamma_j\}}^t(\mathbb{C}_n) \,.
  \label{entprod}
\end{equation}
This formula is equally valid in the non-equilibrium stationary state.

This \emph{ab initio} derivation of the entropy production rate for
non-equilibrium systems with steady mass currents yields results
which are in agreement with the phenomenological prescription of
thermodynamics. Indeed, as we show below, the entropy production rate
of cell $\mathbb{C}_n$, Eq.~(\ref{entprod}), reduces to the phenomenological 
entropy production at corresponding position $X_n$, Eq.~(\ref{thermoep}),
as the grid elements become small and the dependence on 
the choice of partition thus disappears.

\subsection{Multi-baker map}

This formalism is particularly transparent for the multi-baker
map. Referring to Fig.~\ref{fig.mbmap}, and having in mind that measures
are  time-evolved by the inverse map $B_0^{-1}$, it is easy to see that the
bottom horizontal half of cell $n$ is mapped to the left vertical half of
cell $n+1$ and, likewise, the top horizontal half of cell $n$ is mapped to
the right vertical half of cell $n-1$.
Now the invariant measure (\ref{mbmapmuninf}) is uniform with respect to
the $x$ coordinate, which is the expanding direction, and has a fractal
part along the $y$ axis, the contracting direction. Thus the entropy of the
vertical half of cell $n+1$ is half of the entropy of cell $n+1$, but is
not equal to the entropy of the bottom or top horizontal halves of that
cell, at least not with respect to the same partition. 

This observation yields the identification of the entropy production rate,
namely the difference between the entropies of that cell, measured at two
different levels of resolution, one with respect to volumes $d\Gamma = dx
dy$, the other with respect to volumes $d\Gamma' = dx' dy'$, where $dy'$ is
twice as large as $dy$, and $dx'$ half as large as $dx$.

To be precise, given a resolution level $2^{-k}$, $k\in\mathbb{N}$, we
consider the collection of cylinder sets $d\Gamma_k(y_j) = \{(x,y)| y_j\leq
y \leq y_j + 2^{-k}l\}$, $j = 1,\dots,2^k$, with $y_j = 2^{-k}(j-1) l$,
which partition the square into $2^k$ horizontal slabs of widths  
$2^{-k} l$, and, by analogy with Eq.~(\ref{Gibbsentcc}), define the
$k$-entropy of cell $n$ to be 
\begin{equation}
  S_k(\mathbb{C}_n) = - \sum_{j=1}^{2^k}
  \mu_n(d\Gamma_k(y_j)) \left[\log \frac{  \mu_n(d\Gamma_k(y_j))}{2^{-k}}
  - 1\right].
  \label{mbmapkent}
\end{equation}
According to Eq.~(\ref{entprod}), the $k$-entropy production rate per unit
time is
\begin{eqnarray}
  \Delta_\mathrm{i}^\tau S_k(\mathbb{C}_n) &=& \frac{1}{\tau} \left[
    S_k(\mathbb{C}_n)  - S_{k+1}(\mathbb{C}_n) \right],
  \label{mbmapkep}\\
  &=& \sum_{j=1}^{2^{k+1}}
  \mu_n(d\Gamma_{k+1}(y_j)) \log \frac{2 \mu_n(d\Gamma_{k+1}(y_j))}
  {\mu_n(d\Gamma_k(y_{j/2}))}\, ,
  \nonumber
\end{eqnarray}
where the index $j/2$ is the largest integer above $j/2$.

The thermodynamic entropy production rate is recovered in the continuum
limit, whereby the local gradients tend to zero and $k$ can be arbitrarily
large.

To verify this, we consider the symbolic sequences $\uo{k} \equiv
\{\omega_0, \dots, \omega_{k-1}\}$, $\omega_j\in\{0,1\}$ associated to the
$2^k$ points $y_j$ through the dyadic expansions $y_j = y(\uo{k}) =
\sum_{j=0}^{k-1} 2^{-j-1}\omega_j$.

In terms of the symbolic sequences $\uo{k}$, the invariant measure
associated to the cylinder set $d\Gamma(\uo{k}) \equiv d\Gamma(y(\uo{k}))$
becomes, using Eq.~(\ref{mbmapmuninf}),
\begin{eqnarray}
  \mu_n(\uo{k}) &=& 2^{-k}\mu_n + \nabla\rho\Delta T(\uo{k})\,, 
  \nonumber\\
  &=&2^{-k}\mu_n\left[1 + \frac{\nabla\rho}{\mu_n} 
    2^k\Delta T(\uo{k})\right]\,, 
  \label{mukmbmap}
\end{eqnarray}
where we
wrote the local density gradient $\nabla\rho\equiv (\rho_+-\rho_-)l/L$,
and $\Delta T(\uo{k}) = T\big(y(\uo{k})/l + 2^{-k}\big) -
T\big(y(\uo{k})/l\big)$. Using the functional equation (\ref{takagi}), it is
straightforward to check that
\begin{equation}
  2^k \Delta T(\uo{k}) = \sum_{j=0}^{k-1} (1 - 2\omega_j)\,,
  \label{DeltaTk}
\end{equation}
which, up to the scale $l$, is the displacement associated to points in the
cylinder set $d\Gamma(\uo{k})$. Two identities are immediate
\begin{equation}
  \begin{array}{l}
    \sum_{\uo{k}} 2^k \Delta T(\uo{k}) = 0,\\
    \sum_{\uo{k}} 2^k \Delta T(\uo{k})^2 = k.
  \end{array}
  \label{idtakagi}
\end{equation}

Substituting this expression into Eq.~(\ref{mbmapkep}), we arrive, after
expanding in powers of the density gradient, to the expression
\begin{eqnarray}
  \lefteqn{\Delta_\mathrm{i}^\tau S_k(\mathbb{C}_n) =
    \frac{1}{2\tau}\frac{(\nabla\rho)^2}{\mu_n}}\nonumber\\  
  &&\times 
  \left\{ \sum_{\uo{k+1}} 2^{k+1} [\Delta T(\uo{k+1})]^2
    -\sum_{\uo{k}} 2^{k} [\Delta T(\uo{k})]^2 \right\},
  \nonumber\\
  &=& \frac{1}{2\tau}\frac{(\nabla\rho)^2}{\mu_n}
  = \frac{l^2}{2 \tau L^2}\frac{(\rho_+ - \rho_-)^2}{\mu_n}\,,
  \label{mbmapkep2}
\end{eqnarray}
Identifying the diffusion coefficient $\mathcal{D} = l^2 D/\tau$,
with  $D \equiv 1/2$ the diffusion coefficient associated to the binary
random walk, we recover
the phenomenological entropy production rate (\ref{epolg}) for the
macroscopic position variable $X = nl/L$~:
\begin{equation}
  \lim_{\tau, l\to0} \Delta_\mathrm{i}^\tau S_k(\mathbb{C}_n) 
  = \frac{d_\mathrm{i} \mathcal{S}(X_n = nl)}{dt}\,,
  \label{keplimit}
\end{equation}
where the limit $\tau,l \to 0$ is take with the ration $l^2/\tau$ fixed.

\subsection{Lorentz gas}

This computation transposes verbatim to the non-equilibrium stationary
state of the Lorentz gas, Eq.~(\ref{lginfss}), whose entropy production
rate is computed as follows.

Let $\mathbb{M}$ be a partition of $\partial\mathbb{C}$, the phase space of
the collision map, into non-overlapping sets $A_j$, $\mathbb{M} = \cup_j
A_j$. We assume that each element of the partition is contained in a cell
$\partial \mathbb{C}_n$, {\em i.e.} $\forall j\,\exists n : \, A_j
\subset \partial \mathbb{C}_n$. There is thus a sub-collection 
$\mathbb{M}_n$ of cells $A_j$ which partition $\partial \mathbb{C}_n$,
$\partial \mathbb{C}_n = \cup_{j\,:\,A_j\in\mathbb{M}_n} A_j$. Moreover we
assume all the partitions $\mathbb{M}_n$ are isomorphic, which is to
say they can be obtained from one another by translation. 

\begin{figure}[thb]
  \centering
  \includegraphics[angle=0,width=.4\textwidth,height=.4\textwidth]
  {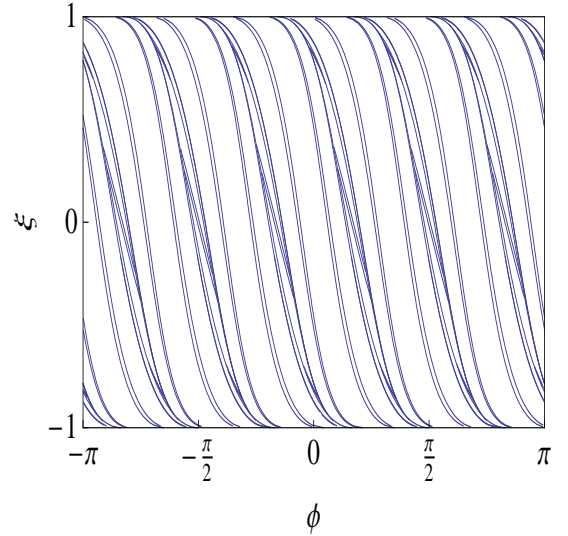}
  \caption{The partition of a unit cell of the open periodic Lorentz
    gas constructed upon two iterations of the collision map. The borders
    of the elements of the partition are the lines of     discontinuity of
    the collision map in the $(\phi,\xi)$ plane. Each point  
    of the curves belongs to a trajectory that grazes a disk in
    one or two iterates.}
  \label{fig.olgpartition}
\end{figure}

We construct such a partition of the phase space associated to cell $n$ by
dividing up $\partial \mathbb{C}_n$ into 
sets $A_j$ according to the displacements of points $\Gamma\in A_j$, such
as shown in Fig. \ref{fig.olgpartition}. Thus a
$k$ partition $\mathbb{M}_n^k$ of $\mathbb{C}_n$ is a collection of 
sets $A(\omega_0,\dots,\omega_{k-1})$ of points $\Gamma$ which have
coherent horizontal displacements for the first $k$ steps. These
displacements are coded by sequences of symbols $\uo{k}$ which, in
this case, take at most $12$ possible values, corresponding to the twelve
possible transitions to the nearest and next-nearest neighboring
disks. However, unlike with Bernoulli processes, not all symbol sequences
are allowed so that the number of sets in the partition $\mathbb{M}_n^k$ is
much less than $12^k$. This is referred to as pruning \cite{CEG92,CEG95}. 

Assuming $\omega_j \in\{1, \dots, 12\}$, we define $a(\omega_j)$ as the
displacement along the horizontal axis corresponding to the label
$\omega_j$, and 
\begin{equation}
  \delta(\uo{k}) \equiv \sum_{j=0}^{k-1}
 a(\omega_j)\,,
  \label{deltak1}
\end{equation}
which, up to the local length scale $l$, is the displacement associated to
the sequence $\uo{k}$, measured with respect to the position of the
scatterers.

The measure associated to these sets is defined by
\begin{equation}
  \mu(\uo{k}) = 
  \int_{A(\uo{k})} d\Gamma \rho(\Gamma)\,,
  \label{lgmuk1}
\end{equation}
where $\rho$ is the invariant density associated to the flux boundary
conditions, Eq.~(\ref{lginfss}). Inserting this expression into
Eq.~(\ref{lgmuk1}), we may write
\begin{eqnarray}
  \lefteqn{
    \mu(\uo{k}) = 
    \mu_n \nu(\uo{k})}
  \label{lgmuk2}\\
  &&+ \frac{\rho_+ - \rho_-}{L}
  \int_{A(\uo{k})} d\Gamma 
  \sum_{k=1}^\infty [x(\Phi^{-t_k}\Gamma) - x(\Phi^{-t_{k-1}}\Gamma)]\,,
  \nonumber
\end{eqnarray}
where
\begin{equation}
  \mu_n = \frac{\rho_++\rho_-}{2} + (\rho_+-\rho_-)\frac{nl}{L} \,,
  \label{lgmun}
\end{equation}
and $\nu(\uo{k}) = \int_{A(\uo{k})} d\Gamma$ is the volume measure
associated to the set $A(\uo{k})$, with $\nu(\uo{k}) =0$ whenever $\uo{k}$
is a pruned sequence. 

We claim that we can make the
approximation 
\begin{eqnarray}
  \frac{1}{\nu(\uo{k})} \int_{A(\uo{k})} d\Gamma 
  \frac{1}{l} \sum_{j=1}^\infty 
  \Big[x(\Phi^{-t_j}\Gamma) &-& x(\Phi^{-t_{j-1}}\Gamma)\Big]
  \nonumber\\
  &\simeq& \delta(\uo{k})\,,  \label{deltak2}
\end{eqnarray}
which we expect to hold provided $k$ is large. This quantity therefore
plays, for the non-equilibrium stationary state of the Lorentz gas, a role
similar to that played under the same conditions by $\Delta
T(\uo{k})/\nu(\uo{k})$, Eq.~(\ref{DeltaTk}), for the multi-baker
map. Indeed the difference 
between the two sides of Eq.~(\ref{deltak2}) is $\mathcal{O}(1)$. 
Under this condition and assuming $l\ll1$, we thus write
\begin{equation}
  \mu(\uo{k}) = \nu(\uo{k})\mu_n \left[1 + \frac{\rho_+ - \rho_-}{\mu_n}
    \frac{l}{L}\delta(\uo{k})\right]\,,
  \label{lgmuk3}
\end{equation}
which can be compared to Eq.~(\ref{mukmbmap}). When $k$ is large,
identities similar to Eq.~(\ref{idtakagi}) hold for $\delta(\uo{k})$~:
\begin{equation}
  \begin{array}{l}
    \sum_{\uo{k}} \nu(\uo{k}) \delta (\uo{k}) = 0\,,\\
    \sum_{\uo{k}} \nu(\uo{k}) \delta (\uo{k})^2 = 2 D k\,,
  \end{array}
  \label{lgidtakagi}
\end{equation}
where the diffusion coefficient $D$ is that of the random walk on half
integer lattice sites associated to the displacements (\ref{deltak1}) with
weights $\nu(\uo{k})$.

The computation of the entropy production rate (\ref{entprod}) then
proceeds along the lines of Eq.~(\ref{mbmapkep2}), with a similar
result: 
\begin{equation}
  \Delta_\mathrm{i}^\tau S_k(\mathbb{C}_n)
  = \frac{l^2 D}{\tau L^2}\frac{(\rho_+ - \rho_-)}{\mu_n}\,,
  \label{lgepk}
\end{equation}
in agreement with the phenomenological expression Eq.~(\ref{epolg}) with
diffusion coefficient $\mathcal{D} = l^2 D/\tau$ and position $X_n =
nl/2$. 

\subsection{Interacting particle system}

Here, as in Ref. \cite{DGG02}, we assume a partition of phase-space in sets
$A_j$ sufficiently small that all points in them have particle No. 1 flowing
through the same sequence of cells over a large time interval, \emph{i.e.}
such that the set of points $\Phi^{-\tau} A_j$ are in the same cell when
projected along the coordinate of particle No. 1, with $0\leq
\tau \leq \mathcal{T}$. The corresponding location is $L(\Phi^{-\tau} \{n,
\Gamma_Q\})$, $\Gamma_Q \in A_j$. We define the displacement
\begin{equation}
  \mathrm{d}(\Gamma_Q, t) = L(\Phi^{-t} \{n, \Gamma_Q\}) - nl\,,
\end{equation}
and write the non-equilibrium stationary state as
\begin{equation}
  \mu(A_j) = \mu_n \nu(A_j) + \frac{\rho_+ - \rho_-}{L}
  \int_{A_j} d\Gamma_Q \mathrm{d}(\Gamma_Q, \mathcal{T})\,,
  \label{spsmua}
\end{equation}
where the volume measure $\nu(A_j)$ is the micro-canonical measure of the
periodic system of $Q$ particles. The integral in this expression encodes
the fluctuating part of the non-equilibrium stationary state. In analogy to
Eq.~(\ref{deltak2}), we can write
\begin{equation}
  \delta(A_j) = \frac{1}{\nu(A_j)} \int_{A_j} 
  d\Gamma_Q \mathrm{d}(\Gamma_Q, \mathcal{T})\,,
\end{equation}
in terms of which we have the identities
\begin{equation}
  \begin{array}{l}
    \sum_{j} \nu(A_j) \delta (A_j) = 0\,,\\
    \sum_{j} \nu(A_j) \delta (A_j)^2 = 2 \mathcal{D} \mathcal{T}\,,
  \end{array}
  \label{spsidtakagi}
\end{equation}
which, apart for their dimensions, are similar to Eqs.~(\ref{idtakagi}) and
(\ref{lgidtakagi}). 

Proceeding as with the multi-baker map and Lorentz gas, we have the entropy
production rate
\begin{equation}
  \Delta_\mathrm{i}^\tau S_\mathcal{T}(\mathbb{C}_n)
  = \frac{\mathcal{D}}{L^2}\frac{(\rho_+ - \rho_-)}{\mu_n}\,,
  \label{lgepk2}
\end{equation}
independently of the resolution scale set by the time parameter
$\mathcal{T}$. 

\section{\label{sec.con}Conclusion}

In this paper we have considered a large class of diffusive deterministic
systems with volume-preserving dynamics, and established the
fractality of the non-equilibrium stationary states which result from the
stochastic injection of particles at the systems' boundaries. 

Under the assumption that the dynamics is chaotic, the natural
non-equilibrium invariant measure is characterized by a phase-space
density which is the stationary state of a Liouville or Perron-Frobenius
operator with flux boundary conditions, which account for 
the presence of particle reservoirs. This density naturally
splits into regular and singular parts. The regular part has a local
equilibrium form, uniform with respect to the scale of macroscopic
variables. It is the natural counterpart on phase space of the solution of
the macroscopic transport equation under non-equilibrium boundary
conditions, displaying a linear density gradient. In contrast, the singular
part varies non-continuously on  
microscopic scales, reflecting the sensitive dependence on initial
conditions, and has no macroscopic counterpart. 

The fundamental achievement of this paper is to have exhibited the
universal features of the non-equilibrium stationary states of simple
low-dimensional deterministic models of diffusion, which are shared by
higher dimensional spatially periodic systems modeling tagged particle
diffusion. As already pointed out in \cite{Gas97}, the fractality of the
non-equilibrium stationary states is the key to a systematic computation of
the positive entropy production rate. Indeed the fractality of phase-space
distributions imposes the use of coarse-graining techniques in order to
define the entropy. The coarse-graining of phase-space into partitions of 
sets of strictly positive volumes results into an entropy which is
generally not constant in time. The decomposition of the time evolution of
such coarse-grained entropies into entropy flux and entropy production
terms provides a framework in which the positiveness of the entropy
production rate can be rigorously established. Furthermore, in the 
macroscopic limit, two key results are obtained: (i) the entropy production
rate is independent of the coarse graining in the limit of arbitrarily fine
phase-space graining, and (ii) its value is  identical to the
phenomenological rate of entropy production, according to thermodynamics.

In this respect, this paper helps clarify a point that had been overlooked
in previous works. Under the diffusive scaling limit, the local particle
density gradient is a natural small parameter in the continuous limit:
fixing the injection rates and system size, we let the spacing between cells
go to zero in order to obtain the 
macroscopic limit. With the identification of this small parameter, it is
straightforward to verify that the computation of the entropy production of
the non-equilibrium stationary states, whether of multi-baker maps, Lorentz
gases, or spatially periodic systems, all yield leading contributions
consistent with the prescription of thermodynamics.

As we have pointed out, the formalism we use relies systematically on the
spatial periodicity of the dynamics, especially with regards to identifying
the rate of entropy production. Though this periodicity is convenient
as it enables one to clearly identify a separation of scales between
microscopic and macroscopic motions, it is a rather restrictive and
ultimately undesirable feature of our models. One would rather describe the
diffusive motion of dilute tracer particles in arbitrary systems. It is our
hope that this work helps set the stage to achieve this more ambitious goal
in future works.

As mentioned in the introduction, this paper is the first of two. In the
companion paper \cite{BGGii}, we consider the influence of an external
field on spatially periodic diffusive volume-preserving systems,
\emph{i.e.} forced systems in 
which no dissipative mechanism is present. As proved by Chernov and
Dolgopyat \cite{CD07a}, such systems are recurrent in the sense that tracer
particles keep coming back to the region of near zero velocity where all
the energy is transferred into potential energy, so that no net current
takes place. We show that these systems 
remain diffusive, albeit with a velocity-dependent diffusion coefficient
which alters the usual scaling laws and consider their statistical
properties in some details. 

\begin{acknowledgments}
This research is financially supported by the Belgian Federal 
Government  (IAP project ``NOSY") and the ``Communaut\'e fran\c caise de
Belgique'' (contract ``Actions de Recherche Concert\'ees''
No. 04/09-312) as well as by the Chilean Fondecyt under 
International Cooperation Project 7070289. 
TG is financially supported by the Fonds de la Recherche
Scientifique F.R.S.-FNRS. FB acknowledges financial support from the 
Fondecyt Project 1060820 and FONDAP 11980002 and Anillo ACT 15. 
\end{acknowledgments}

\end{document}